\documentclass[12pt]{article}

\usepackage{amsmath,amssymb,epsfig}
\usepackage[all]{xypic}

\usepackage{color}
\definecolor{darkgreen}{cmyk}{1,0,1,.2}
\definecolor{m}{rgb}{1,0.1,1}

\newcommand{\be}{\begin{equation}}
\newcommand{\ba}{\begin{eqnarray}}
\newcommand{\ea}{\end{eqnarray}}

\def\ca{{\cal A}}

\newcommand{\bbC}{{\Bbb C}}

\newcommand{\bbR}{{\Bbb R}}

\fontfamily{yfrak}

\begin{document}

\vskip 15mm

\begin{center}

{\Large\bfseries Lattice Loop Quantum Gravity
%\\[2mm]
}

\vskip 4ex

Johannes \textsc{Aastrup}$\,^{a}$\footnote{email: \texttt{johannes.aastrup@uni-muenster.de}} \&
Jesper M\o ller \textsc{Grimstrup}\,$^{b}$\footnote{email: \texttt{grimstrup@nbi.dk}}

\vskip 3ex  

$^{a}\,$\textit{Mathematical Institute, University of M\"unster,\\ Einsteinstrasse 62, D-48149 M\"unster, Germany}
\\[3ex]
$^{b}\,$\textit{The Niels Bohr Institute, University of Copenhagen, \\Blegdamsvej 17, DK-2100 Copenhagen, Denmark}

\end{center}

\vskip 3ex

\begin{abstract}
We present a separable version of Loop Quantum Gravity (LQG) based on an inductive system of cubic lattices. We construct semi-classical states for which the LQG operators -- the flux, the area and the volume operators -- have the right classical limits. Also, we present the Hamilton and diffeomorphism constraints as operator constraints and show that they have the right classical limit. Finally, we speculate whether the continuum limit, which these semi-classical states probe, can be defined for the entire construction and thereby restore an action of the diffeomorphism group.

\end{abstract}

\newpage

\section{Introduction}
A critical challenge for the program of Loop Quantum Gravity (LQG) is the formulation of a semi-classical limit that reproduces General Relativity (for a survey on LQG see \cite{AL1}).
%Here, a key issue seems to be separability of the relevant Hilbert space to secure normalizability of the semi-classical states. 
In this note we propose a separable version of LQG, based on an ordered system of nested graphs, and show that it contains normalizable semi-classical states for
which the LQG operators, including a version of the Hamiltonian, descents to their classical counterparts. 

Essentially, the model which we present is identical to the traditional formulation of LQG except that the kinematical Hilbert space is constructed as an inductive limit taken over an infinite set of 3-dimensional, nested cubic lattices. These lattices have a natural interpretation as the coordinate system in which the Ashtekar variables are formulated. On a classical level the holonomy and flux functions living in a system of cubical lattices separates the points in the phase space, and therefore a rich enough system of functions to completely describe the classical system.

This formulation of LQG is not a priori diffeomorphism invariant, since it explicitly depends on a chosen coordinate system, nor is it invariant under rotations and translations. Worse yet, it does not even permit an action of these symmetry groups and thus this approach may seem questionable. However, results on semi-classical states obtained in the article  \cite{AGNP1} have convinced us that this approach deserves further analysis.

The article \cite{AGNP1} deals with the physical interpretation of spectral triples constructed over a configuration space of connections closely related to the configuration space in LQG, see \cite{AGN1,AGN2, AGN3}. Essentially, this spectral triple is the construction of a Dirac type operator {\it on} the space of connections. The main achievement of the article was the construction of semi-classical states on which this operator reduces to the Dirac Hamiltonian of a fermion coupled to gravity in 3+1 dimensions. The construction of this limit apparently requires a system of nested lattices. 

The insight obtained in \cite{AGNP1} is that it seems possible, from a setup based on nested, cubic lattices, to restore all spatial symmetries in a kind of continuum limit. This limit is constructed so that all dependency on finite parts of the system of lattices is removed. We believe that this points towards a general method of quantization, where the lattices only serve as an intermediate step and where the final continuum limit discards dependencies on finite parts of these lattices. This resembles the continuum limit of lattice gauge theory and is analog to the way that a Riemannian integral does not depend on quantities of measure zero. 

Presently we only know how to perform this continuum limit for specific states and only in combination with a classical approximation. We suspect, however, that it may be possible to perform this limit for the whole construction, that is, for the algebra and for the Hilbert space. %Therefore, we believe that it makes sense to consider a formulation of LQG based on cubic lattices.

This paper differs from \cite{AGNP1} in that it does not deal with a construction of an Euler-Dirac type operator, but only with the quantized operators of LQG, i.e. the holonomy operators, the flux operators, the area operators and the volume operators, as well as a version of a Hamilton operator. As a consequence the involved Hilbert space is the space of square integrable functions of a completed space of connections, and tensoring with the CAR-algebra is not required.   
We consider the semi-classical states constructed in \cite{AGNP1}, adopted to the setup without the CAR-algebra, and show that all the LQG-operators have the right semi-classical limits.\\

This paper is organized as follows: In section 2 we briefly review the classical setup, General Relativity formulated in terms of connection variables, as well as the holonomy and flux variables. In section 3 we construct the kinematical Hilbert space and represent the holonomy and flux variables as operators hereon. Section 4 deals the semi-classical states and in section 5 and section 6 we show that the flux, area and volume operators have the right classical limit, as does an operator version of the Hamiltonian. In section 7 we give a discussion.

\section{The connection formalism of gravity}
We will in this section recall the formulation of canonical gravity in terms of connection variables (for details see \cite{AL1}). 

First assume that space-time $M$ is globally hyperbolic. Then $M$ can be foliated as 
$$M=\Sigma \times \bbR, $$   
where $\Sigma$ is a three dimensional hyper surface. We will assume that $\Sigma$ is oriented and compact.

The fields in which we will describe gravity are 
\begin{itemize}
\item $SU(2)$-connections in the trivial bundle over $\Sigma$. These will be denoted $A_i^a$.
\item $\mathfrak{su}(2)$-valued vector densities on $\Sigma$. We will adopt the notation $E_a^i$, where $a$ is the $\mathfrak{su}(2)$-index.  
\end{itemize} 
On the space of field configurations, which we denote $\mathcal{P}$, there is a Poisson bracket expressed in local coordinates by
$$\{ A_i^a (x), E_b^j(y) \} =\delta_i^j \delta_b^a \delta (x,y), $$
where $\delta (x,y)$ is the delta function on $\Sigma$. The rest of the brackets are zero.
These fields are subjected to constraints (euclidian signature) given by
\begin{eqnarray*}
 \epsilon_c^{ab} E^i_a E^j_b F_{ij}^c&=&0\\
  N^i E^j_a F^a_{ij}&=&0    \\
  (\partial_i E^i_a+\epsilon_{ab}^cA_i^b E^i_c)&=&0
\end{eqnarray*}
Here $F$ is the field strength tensor, of the connection $A$. The first constraint is the Hamilton constraint, the second is the diffeomorphism constraint and the third is the Gauss constraint.

These field configurations together with the constraints constitute an equivalent formulation of General Relativity without matter. 

\subsection{Reformulation in terms of holonomy and fluxes}
The crucial point for quantization of the kinematical part of gravity in LQG is the reformulation of the Poisson bracket in terms of holonomies and fluxes. For a given path $p$ in $\Sigma$ the holonomy function is simply the parallel transport along the the path, i.e.
$$\mathcal{P}\ni (A,E) \to Hol (p,A)\in G.$$
Given an oriented surface $S$ in $\Sigma$ the associated flux function is given by
$$\mathcal{P}\ni (A,E)\to \int_S \epsilon_{ijk} E_a^idx^jdx^k.$$
The holonomy function for a path will also be denoted with $h_p$ and the flux function will be denoted by $F^S_a$.

Let $p$ be a path and $S$ be an oriented surface in $\Sigma$ and assume $p$ ends in $S$ and has exactly one intersection point with $S$. The Poisson bracket in this case becomes
\begin{equation}
\{ h_p, F^S_a \}(A,E)=\pm \frac{1}{2} h_p(A)\sigma_a \label{poisson1}
\end{equation} 
where $\sigma_a$ is the Pauli matrix with index $a$. The sign in (\ref{poisson1}) is negative if the orientation of $p$ and $S$ is the same as the orientation of $\Sigma$, and positive if not. If $p$ instead starts in on $S$ one gets the Poisson bracket
\begin{equation}
\{ h_p, F^S_a \}(A,E)=\pm \frac{1}{2}\sigma_a h_p(A) \label{poisson2}
\end{equation}
but now with the reverse sign convention.
If $p$ is contained in $S$ the Poisson bracket is zero.

\section{The kinematical Hilbert space and quantization of the Poisson bracket}
We will in this section give a brief description of the involved Hilbert space and the quantized operators on it. For further details we refer to \cite{AGN3,AGNP1}.

Let $\Gamma_0$ be a cubical lattice on $\Sigma$. We will assume that edges are directed and that these give rise to a coordinate system in such a way that the edges correspond to one unit at the coordinate axis and that the orientation of the coordinate axis coincide with the orientation of $\Sigma$. 

 Let $\Gamma_n$ be $\Gamma_0$ subdivided $n$-times. We define 
$$\ca_n=G^{e(\Gamma_n)},$$
where $e(\Gamma_n)$ is the number of edges in $\Gamma_n$, and $G=SU(2)$. We have hence to each edge in $\Gamma_n$ associated a copy of $G$. A smooth connection $A$ gives rise to a point in $\ca_n$ via
$$\ca \ni \nabla \to (Hol ( A, e_i))_{e_i\hbox{ edge in }\Gamma_n}.$$
There are canonical maps 
$$P_{n+1,n} : \ca_{n+1} \to \ca_n,$$
which simply consist in multiplying the two elements in $G$ attached to an edge in $\Gamma_n$ which gets subdivided in $\Gamma_{n+1}$. Define
$$\overline{\ca}=\lim_n \ca_n,$$
where the limit on the right hand side is the projective limit. Since each $\ca_n$ is a compact topological Haussdorf space, $\overline{\ca}$ is a compact Haussdorf space. It is easy to see that the maps $\ca \to \ca_n$ induces a map
$$\ca \to \overline{\ca},$$
and that this is a dense embedding, see \cite{AGN2,AGN3} 

The main technical tool in \cite{AGN2,AGN3} was that this projective system can be rewritten into a system of the form  
$$G^{e(\Gamma_0)}\leftarrow G^{e(\Gamma_1)}\leftarrow \cdots ,$$
where the maps consist in deleting copies of $G$'s. The way it is rewritten is however not unique. It depends on a labeling of the new degrees of freedoms which appears from going from level $n$ to level $n+1$. We choose to label the edge appearing to the left in a subdivision as the degree of freedom, see figure 1.

\begin{figure}[t]
\begin{center} 
\resizebox{!}{4cm}{
 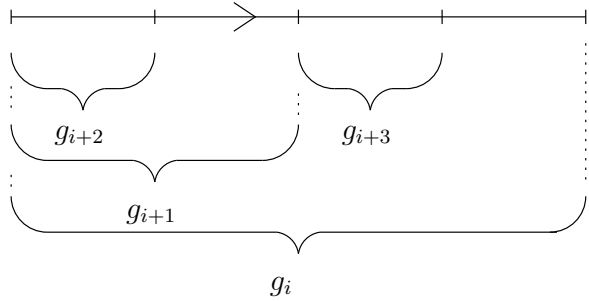}
\end{center}
\caption{\it The chosen labeling.}
\end{figure}

\subsection{The Hilbert space and the flux operators}
Define
$$L^2(\ca_{\Gamma_n})=L^2(G^{e(\Gamma_n)}),$$
where the measure on the right hand side is the normalized Haar measure. Next define
$$L^2(\overline{\ca})=\lim_nL^2 (\ca_{\Gamma_n}).$$
This will be the Hilbert space on which we will define the quantized operators.

A path $p$ in $\cup_n \Gamma_n$  gives rise to a bounded function $h_l$ with values in $M_2$ via
$$\overline{\ca}\ni A \to Hol(A ,p),$$
where $Hol( A ,p)$ is the extension of the holonomy map from $\ca$ to $\overline{\ca}$, see f.ex. \cite{AGN3}. Therefore $h_p$ has a natural action on the Hilbert space $L^2(\overline{\ca})\otimes M_2$.
\begin{figure}[t]
\begin{center} 
\resizebox{!}{5cm}{
 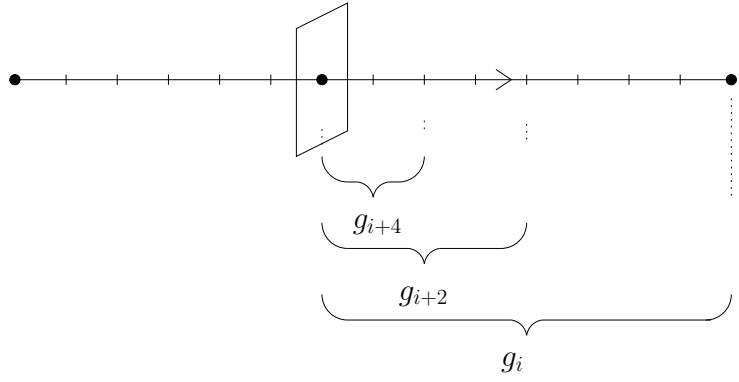}
\end{center}
\caption{\it The copies of $SU(2)$ corresponding to edges with left endpoints coinciding with the right endpoint of $e$. }
\end{figure}
To construct the flux operators first look at an edge $e\in \Gamma_n \setminus \Gamma_{n-1}$. The first guess for a flux operator associated to the infinitesimal surface $S_e$ sitting at the right end point of $e$ is
$$\hat{F}^{S_e}_a=i\frac{1}{2}\left( \mathcal{L}_a+\sum_k \mathcal{R}_a^n\right) ,$$
where $\mathcal{L}_a$ is the left invariant vector field on the copy of $SU(2)$ associated to $e$ corresponding to the generator in $\mathfrak{su}(2)$ with index $a$, and where $\mathcal{R}_a^n$ is the right invariant vector field on the copy of $SU(2)$ corresponding to edge in $k$'th subdivision with left end point in the right endpoint of $e$, see figure 2.

If $p$ is a path which which runs through $e$ and only has one intersection point with $S_e$ then
$$[ h_p,\hat{F}_a^{S_e} ]=i\frac{1}{2}h_p\sigma_a,$$
and if $p$ is a path that leaves $S_e$ on the other side with one intersection point then 
$$[ h_p,\hat{F}_a^{S_e} ]=i\frac{1}{2}\sigma_a h_p,$$
i.e. the relations (\ref{poisson1}),(\ref{poisson2}) are realized as commutator relations.

If  $p$ is a path that runs through $S_e$ and does not involve the copy of $SU(2)$ corresponding to $e$, see figure 3, then 
$$[h_p, \hat{F}^{S_e}_a ]=0.$$

\begin{figure}[t]
\begin{center} 
\resizebox{!}{3cm}{
 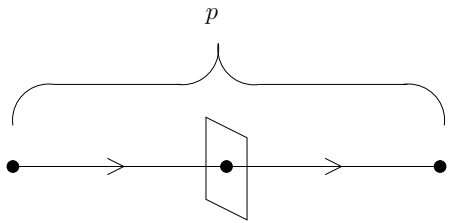}
\end{center}
\caption{\it If $p$ corresponds to a copy of $SU(2)$ it is not acted upon by $\hat{F}^{S_e}_a$. }
\end{figure}

We can remedy this by adding a vector field $\mathcal{O}_{n-1}$, which only acts on copies of $SU(2)$ coming from edges in $\Gamma_{n-1}$, see \cite{AGNP1} for details. Thus
 $$\hat{F}^{S_e}_a=i\frac{1}{2}\left( \mathcal{L}_a+\sum_k \mathcal{R}_a^n\right)+i\mathcal{O}_{n-1} $$
realizes the Poisson bracket (\ref{poisson1}), (\ref{poisson2}).

 %If $p$ is a path which runs one time through $S_e$ and involves the variable corresponding to $e$ it is an easy computation that
%$$[h_p,\hat{F}^{S_e}_a ]=h_{p_1}\sigma_a h_{p_2} ,$$
%where $p_1,p_2$ are the edges apearing from letting $S_e$ divide $p$, see figure?.
%Thus the $\hat{F}_{S_e}^a$ and $h_p$ is a quantization of the Poisson bracket (\ref{poisson1}),(\ref{poisson2}). 

%If the loop does not involve the variable corresponding to $e$, see for example figure ?, the commutator will not give the right expression. It is however easy to see, ..., that by modifying the expression of $\hat{F}_{S_e}^a$ with certain vector fields living on edges in $\Gamma_{n-1}$ one gets the correct commutator. All together we get operators of the form
%$$\hat{F}_{S_e}^a=\mathcal{L}_a+\Omega_{n-1}.$$

\section{The semi-classical states}
We will in this section briefly recall part of the properties of the semi-classical states constructed in \cite{AGNP1}. This construction used results of Hall \cite{H1,H2}, and was inspired by the articles \cite{BT1,TW,BT2}.

Let $E,A$ be a point in the classical phase space. The semi-classical state $\phi_n^t \in L^2(\overline{\ca})$ with respect to this point have the properties:
\begin{enumerate}
\item For any path $p \in \Gamma_n$ and any $w\in M_2(\bbC )$ \label{hol}
$$\lim_{t \to 0}\langle \phi_n^t \otimes w,h_p \phi_n^t\otimes w\rangle= \langle w,Hol(p,A)w\rangle$$
This in particular means that the expectation value in the limit $n\to \infty$ on a path in $\cup_n\Gamma_n$ is just the holonomy the connection $A$ along the path.
\item \label{efelt} For an edge $e\in \Gamma_n \setminus \Gamma_{n-1}$ in direction $i$
$$\lim_{t \to 0}\langle \phi_n^t,t i \mathcal{L}^e_a\phi_n^t\rangle =2^{-2n}E_a^i (v_e),$$  
 where $v_e$ denotes the right end point of $e$. If $e \notin \Gamma_n \setminus \Gamma_{n-1}$ the corresponding expectation value will be zero.
\item $\| \phi_n^t\|=1$.
\end{enumerate}
The properties \ref{hol}.,\ref{efelt}. also hold for polynomial function on $T^*SU(2)$, i.e.
\begin{eqnarray*}
\lefteqn{\lim_{t \to 0}\langle \phi_n^t \otimes w,P(h_l,t i \mathcal{L}^e_1,t i \mathcal{L}^e_2 ,it \mathcal{L}^e_3) \phi_n^t\otimes w\rangle}\\
&=& \langle w,P(Hol(l,A),E_1^i, E_2^i,E_3^i),w\rangle,
\end{eqnarray*}
and for more general functions in $T^*SU(2)$.

Property \ref{hol}. is a consequence of the peakedness of $\phi_n^t$ around $Hol (p,A)$. In particular when the edge becomes small, $\phi_n^t$ is centered around "$1+\epsilon A$". Since left and round invariant vector fields coincide in the identity on the group, for small edges we have
  $$\lim_{t \to 0}\langle \phi_n^t,t i \mathcal{R}^e_a\phi_n^t\rangle \sim 2^{-2n}E_a^i (v_e).$$

\section{The flux, area and volume operator}
In this section we will compute the expectation value of the flux, area and volume operator on the semi-classical states. The definitions of the operators are adoptions of corresponding operators in LQG to the setting of the lattice, in which we are working.

\subsection{The flux operator}
Let $S$ be a surface in $\Sigma$ which consists of parts of faces of the lattice $\cup \Gamma_n$. Define the flux operator
$$\hat{F}^S_a=\frac{8}{7}\sum_{e\in S} \hat{F}^{S_e}_a,$$
where $e \in S$ means that the right end point of $e$ belongs to a part of $S$ orthogonal to $e$. 

Let us for notationally simplicity assume that $S$ is contained in a $x_2,x_3$-plane and with the same orientation. The expectation value gives   
\begin{eqnarray*}
\lefteqn{\lim_{n\to \infty}\lim_{t \to 0} \langle \phi_n^t ,\hat{F}_a^S\phi_n^t\rangle}\\
&= & \frac{8}{7} \lim_{n \to \infty} \left( \sum_{e\in S, e\in \Gamma_n \setminus \Gamma_{n-1}} 2^{-2n}E_1^a(v_e)+ \sum_{e  \in S , e\in \Gamma_{n-1}}2^{-2(n+1)}E_1^a(v_e) \right) \\
&=&\int_S E_a^1dx^2dx^3=F^S_a.
\end{eqnarray*}
Thus expectation value of the flux operator on the semi-classical state reproduces the classical value of the flux operator.

We have in the above calculation used that the expectation value for the right invariant vector fields approaches those of the left invariant vector fields as $n\to \infty$, and we have also used that for edges in $\Gamma_{n-1}$ only the right invariant part of $\hat{F}^{S_e}_a$ contributes, and therefore only with a factor $\frac{1}{2}$. This is the reason the $\frac{8}{7}$ prefactor shows up.

\subsection{The area operator}
The area operator for a surface as above is defined by
$$\hat{A}_S=\frac{8}{7}\sum_{e \in S} \sqrt{\sum_a (F^{S_e}_a)^2}$$

Lets again for notationally simplicity assume that $S$ is contained in a $x_2,x_3$-plane and with the same orientation. The expectation value becomes via the same calculation as for the flux operator
\begin{eqnarray*}
\lim_n \lim_t \langle \phi_n^t, \hat{A}\phi_n^t \rangle =\int_S \sqrt{\sum_a(E_a^1)^2}dx^2dx^3 =A(S),
\end{eqnarray*}
where $A(S)$ is the area of $S$ with respect to the metric given by the field $E$, see for example \cite{AL1}.
\subsection{The volume operator}
Let $v$ be a vertex in end let $e_1,e_2,e_3$ be the three edges running into it which are labeled be a copy of $SU(2)$. We will for simplicity denote $\hat{F}_a^{S_{e_i}}$ by $\hat{F}_a^i$. Define 
$$\hat{V}_v=\sqrt{ |\det \hat{F}^i_a| }.$$
For a space region $R$ in $\Sigma$ define
$$\hat{V}_R=\frac{16 \sqrt{2}}{8 \sqrt{2}+7}\sum_{v\in  R}\hat{V}_v,$$
where $v$ are the vertices in $\cup \Gamma_n$. 

A computation similar to the ones before give
\begin{eqnarray*}
\lefteqn{\lim_n \lim_t\langle \phi_n^t ,\hat{V}_R\phi_n^t\rangle}\\
&=&\lim_{n\to \infty} \sum_{v\in R\cap \Gamma_n}2^{-3n}\sqrt{| \det E_a^i(v) |}=\int_R dx^3\sqrt{| \det E_a^i|}=V(R)
\end{eqnarray*}
where $V(R)$ denotes the volume of the region $R$ with respect to the metric induced by $E$.
\section{A Hamilton constraint}
We will start be rewriting the Hamilton plus the diffeomorphism constraint. Let $N$ and $N^i$ be the lapse and shift field. Then
\begin{eqnarray*}
\lefteqn{N \epsilon_c^{ab}E_a^iE_b^jF^c_{ij}+N^iE_a^jF_{ij}^a}\\
&=& Tr(E_a^i (i\sigma^a) E_b^j (i\sigma^b) F_{ij}^c (i\sigma_c) (N+N^di\sigma_d))
\end{eqnarray*}
where $N^dE_d^i=N^i$, and $Tr$ is the $SU(2)$-trace.

We will give a quantized version of this expression integrated. 

Let $v$ be a vertex in $\Gamma_n\setminus \Gamma_{n-1}$. Let $l_{ij}^n$ be the smallest loop in $\Gamma_n$ in the $i,j$-direction with $v$ as base point.   
Let $e_i, e_j$ be the two vertices in $\Gamma_n\setminus \Gamma_{n-1}$ in the $i,j$ directions running into $v$. Define 
\begin{equation}\label{locham}
\hat{H}_v=2^{3n} Tr(\hat{F}_a^i (i\sigma^a) \hat{F}_b^j (i\sigma^b) i(l_{ij}^n-1)  (N+N^di\sigma_d)),
\end{equation}

Finally define
$$\hat{H}=\sum_v\left( \frac{8}{7}\right)^2 \hat{H}_v,$$
where the sum runs over all vertices with property above and all $n$.

Using that 
$$\lim_t \langle \phi_n^t\otimes w ,l_{ij}^n \phi_n^t \otimes w \rangle \sim \langle w ,(1+2^{-2n}F_{ij})w\rangle $$
 when $n$ approaches infinity we get
\begin{eqnarray*}
\lefteqn{\lim_n \lim_t \langle \phi_t^n,\hat{H} \phi_t^n\rangle} \\
&=& \int_\Sigma dx^3 Tr(E_a^i (i\sigma^a) E_b^j (i\sigma^b) F_{ij}^c (i\sigma_c) (N+N^di\sigma_d)) 
\end{eqnarray*}
The prefactor $2^{3n}$ in (\ref{locham}) shows up because we are quantizing the Hamilton constraint without the inverse determinant. 

\section{Discussion}

The crucial difference between the approach proposed in this paper and that of LQG is the absence of the action of the diffeomorphism group. To justify the approach proposed here one must identify a mechanism which restores such an action. The construction of the semi-classical states hints, in our opinion, towards such a mechanism, since it shows that the symmetries are restored in a certain continuum limit combined with a classical limit. The idea is to understand the entire theory as a continuum limes, much alike lattice gauge theory, where any dependency on finite parts of the lattice vanishes. The difference between this approach and lattice gauge theory, then, is that this continuum limit is not a priori dependent on a background metric.

To be concrete, what we have in mind is to take the continuum limit without the semi-classical limit. This means that the states which we aim to consider 'live' on infinitely small edges and varies continuously in the space. Thus, we wish to consider sequences of states, labelled by depth in the inductive system of lattices, and to impose a kind of smoothness condition on these states with respect to the underlying manifold.

A technical issue is the question of convergence. In this paper we have computed the $\lim_n \lim_t$ limit, whereas the appropriate thing to do would be to prove that the limit as $n \to \infty$ exists, and then take the classical limit $t \to 0$. This will of course require further analysis.      

Notice that the type of semi-classical states described in this paper is not a complete set of states. The reason being that they are insensitive to anything living on finite parts of the lattices. This corresponds to a large amount of arbitrariness concerning the operators with respect to this continuum limit. This should, in our opinion, again indicate a limit of the entire construction so that such semi-classical states indeed form a complete set in some limiting Hilbert space, and where operators are defined uniquely.

So far, the construction is set with a Euclidean signature. The Lorentzian case is characterized by a complexified $SU(2)$ connection. In the LQG setup, the Lorentzian case is problematic since it corresponds to a non-compact group which obstructs the construction of the kinematical Hilbert space. Briefly stated, the inductive limit of Hilbert spaces requires the identity function on the group to be square integrable, something which requires the space to be compact. However in the continuum limit which we have introduced one works with connections and not the holonomies, i.e. with the Lie algebra and not the group itself. Thus, one could adopt the strategy to work with $SU(2)$ and then obtain a complex connection by doubling the Hilbert space, inserting a complex $i$ in appropriate places and take the continuum limit.

In section 6 we have written down the Hamilton and Diffeomorphism constraint as a single operator constraint. Thus, we aim at a construction which treats these constraints on an equal footing.
Concerning the Hamiltonian we should point out that the expression which we write in section 6 depends strongly on the chosen coordinate system since it lacks a factor with the inverse of the square root of the determinant of the metric. Thus, this expression should be understood strictly in terms of a constraint equation.  Alternatively, one may add this missing factor. However, this seems to ruin the otherwise aesthetically attractive form of the constraints. Furthermore, we find it interesting that this issue introduces a scaling degree of freedom in the construction. One may speculate whether this points towards a connections with renormalization group theory.

%Diskussion: Ordinary  Separation of E-fields, $n,t$-convergence, Gauge equivalence, Hilbert space, diffeomorphisms, continuum limit, euclidian vs lorentz, compact vs noncompact, lack of uniqueness, inverse volume, constaints vs. Hamilton. Dif constraint as operator constraint. Renormalization, scale.

\end{document}